\documentstyle[12pt]{article}

\def\frac#1#2{{#1 \over #2}}

\begin{document}
\setcounter{page}{0}

\newcommand{\R}{{\rm I\!R}}
\newcommand{\Z}{{\sf Z\!\!Z}}
\newcommand{\goto}{\rightarrow}
\newcommand{\Goto}{\Rightarrow}

\newcommand{\al}{\alpha}
\renewcommand{\b}{\beta}
\renewcommand{\c}{\chi}
\renewcommand{\d}{\delta}
\newcommand{\D}{\Delta}
\newcommand{\ve}{\varepsilon}
\newcommand{\f}{\phi}
\newcommand{\F}{\Phi}
\newcommand{\vf}{\varphi}
\newcommand{\g}{\gamma}
\newcommand{\ii}{\iota}
\newcommand{\G}{\Gamma}
\newcommand{\k}{\kappa}
\renewcommand{\l}{\lambda}
\renewcommand{\L}{\Lambda}
\newcommand{\m}{\mu}
\newcommand{\n}{\nu}
\newcommand{\r}{\rho}
\newcommand{\vr}{\varrho}
\renewcommand{\o}{\omega}
\renewcommand{\O}{\Omega}
\newcommand{\p}{\psi}
\renewcommand{\P}{\Psi}
\newcommand{\s}{\sigma}
\renewcommand{\S}{\Sigma}
\newcommand{\th}{\theta}
\newcommand{\vt}{\vartheta}
\renewcommand{\t}{\tau}
\newcommand{\vp}{\varpi}
\newcommand{\x}{\xi}
\newcommand{\z}{\zeta}
\newcommand{\ta}{\triangle}
\newcommand{\w}{\wedge}
\newcommand{\e}{\eta}
\newcommand{\Th}{\Theta}
\newcommand{\td}{\tilde}
\newcommand{\ep}{\epsilon}
\newcommand{\na}{\nabla}
\newcommand{\be}{\begin{equation}}
\newcommand{\ee}{\end{equation}}
\newcommand{\eE}[1]{\label{#1}\end{equation}}
\newcommand{\Be}{\begin{eqnarray}}
\newcommand{\Ee}{\end{eqnarray}}
\newcommand{\EE}[1]{\label{#1}\end{eqnarray}}
\newcommand{\ba}{\begin{array}}
\newcommand{\bi}{\begin{itemize}}
\newcommand{\ei}{\end{itemize}}
\newcommand{\ea}{\end{array}}
\newcommand{\nn}{\nonumber}
\newcommand{\wt}{\widetilde}
\renewcommand{\thefootnote}{\fnsymbol{footnote}}

\newcommand{\ind}{\scriptscriptstyle}

\newcommand{\CAG}{{\cal A/\cal G}}
\newcommand{\CA}{{\cal A}}
\newcommand{\CB}{{\cal B}}
\newcommand{\CC}{{\cal C}}
\newcommand{\CD}{{\cal D}}
\newcommand{\CE}{{\cal E}}
\newcommand{\CF}{{\cal F}}
\newcommand{\CG}{{\cal G}}
\newcommand{\CH}{{\cal H}}
\newcommand{\CI}{{\cal I}}
\newcommand{\CL}{{\cal L}}
\newcommand{\CO}{{\cal O}}
\newcommand{\CP}{{\cal P}}
\newcommand{\CQ}{{\cal Q}}
\newcommand{\CR}{{\cal R}}
\newcommand{\CT}{{\cal T}}
\newcommand{\CU}{{\cal U}}
\newcommand{\CW}{{\cal W}}
\newcommand{\CM}{{\cal M}}
\newcommand{\CS}{{\cal S}}
\newcommand{\CN}{{\cal N}}

\newcommand{\rd}{\partial}
\newcommand{\grad}[1]{\,\nabla\!_{{#1}}\,}
\newcommand{\gradd}[2]{\,\nabla\!_{{#1}}\nabla\!_{{#2}}\,}
\newcommand{\om}[2]{\omega^{#1}{}_{#2}}
\newcommand{\vev}[1]{\langle #1 \rangle}
\newcommand{\lrarr}{\longrightarrow}
\newcommand{\darr}[1]{\raise1.5ex\hbox{$\leftrightarrow$}\mkern-16.5mu #1}
\newcommand{\Ha}{{\frac{1}{2}}}
\newcommand{\ha}{{\textstyle{\frac{1}{2}}}}
\newcommand{\fr}[2]{{\textstyle{#1\over#2}}}
\newcommand{\Fr}[2]{{#1 \over #2}}
\newcommand{\dt}{{\frac{d}{dt}}}
\newcommand{\rf}[1]{\fr{\rd}{\rd #1}}
\newcommand{\rF}[1]{\Fr{\rd}{\rd #1}}
\newcommand{\Rf}[2]{\fr{\rd #1}{\rd #2}}
\newcommand{\RF}[2]{\Fr{\rd #1}{\rd #2}}
\newcommand{\df}[1]{\fr{\d}{\d #1}}
\newcommand{\dF}[1]{\Fr{\d}{\d #1}}
\newcommand{\DF}[2]{\Fr{\d #1}{\d #2}}
\newcommand{\DDF}[3]{\Fr{\d^2 #1}{\d #2\d #3}}
\newcommand{\ddf}[2]{\Fr{d^2 #1}{d #2^2}}
\newcommand{\DDDF}[4]{\Fr{\d^3 #1}{\d #2\d #3\d #4}}
\newcommand{\ddF}[3]{\Fr{\d^n#1}{\d#2\cdots\d#3}}
\newcommand{\fs}[1]{#1\!\!\!/\,}   
\newcommand{\Fs}[1]{#1\!\!\!\!/\,} 
\newcommand{\pa}{p_a \dot{x}^a}
\newcommand{\slag}{\Fr{i}{4}\l ^{ab}\,Tr\,\s_{ab}\,\L^{-1}\dot{\L}}
\newcommand{\lag}{\CL\,=p_a \dot{z}^a+\Fr{i}{2}\l ^{ab}\,Tr,s_{ab}\
\L^{-1}\dot{\L}}
\newcommand{\spin}{\Fr{1}{2}\l^{ab}\L\,\s_{ab}\,\L^{-1}}
\newcommand{\Df}[1]{\Fr{d}{d #1}}
\newcommand{\hlkl}{\Fr{\l^{kl}}{2}}
\newcommand{\stcon}{C_{ab,cd}{}{}{}^{ef}}

\newcommand{\cmp}[3]{, Comm.\ Math.\ Phys.\ {{\bf #1}} {(#2)} {#3}.}
\newcommand{\pl}[3]{, Phys.\ Lett.\ {{\bf #1}} {(#2)} {#3}.}
\newcommand{\np}[3]{, Nucl.\ Phys.\ {{\bf #1}} {(#2)} {#3}.}
\newcommand{\pr}[3]{, Phys.\ Rev.\ {{\bf #1}} {(#2)} {#3}.}
\newcommand{\prl}[3]{, Phys.\ Rev.\ Lett.\ {{\bf #1}} {(#2)} {#3}.}
\newcommand{\ijmp}[3]{, Int.\ J.\ Mod.\ Phys.\ {{\bf #1}} {(#2)} {#3}.}
\newcommand{\mpl}[3]{, Mod.\ Phys.\ Lett.\ {{\bf #1}} {(#2)} {#3}.}
\newcommand{\jdg}[3]{, J.\ Diff.\ Geo.\ {{\bf #1}} {(#2)} {#3}.}
\newcommand{\pnas}[3]{, Proc.\ Nat.\ Acad.\ Sci.\ USA.\ {{\bf #1}} {(#2)}
{#3}.}
\newcommand{\zp}[3]{, Z.\ Phys.\ {{\bf #1}} {(#2)} {#3}.}
\newcommand{\ap}[3]{, Ann.\ Phys.\ {{\bf #1}} {(#2)} {#3}.}
\newcommand{\jmp}[3]{, J.\ Math.\ Phys.\ {{\bf #1}} {(#2)} {#3}.}
\newcommand{\nc}[3]{, Il\ Nuovo\ Cimento\ {{\bf #1}} {(#2)} {#3}.}
\newcommand{\prs}[3]{, Proc.\ Roy.\ Soc.(London)\ {{\bf #1}} {(#2)} {#3}.}
\newcommand{\rmp}[3]{, Rev.\ Mod.\ Phys.\ {{\bf #1}} {(#2)} {#3}.}
\newcommand{\grg}[3]{, Gen.\ Rel.\ Grav.\ {{\bf #1}} {(#2)} {#3}.}
\newcommand{\cqg}[3]{, Class.\ Quantum \ Grav.\ {{\bf #1}} {(#2)} {#3}.}
\newpage
\setcounter{page}{0}
\begin{titlepage}
\begin{flushright}
\hfill{KAIST-CHEP-94/20}\\
\hfill{SNUTP 94-80}\\
\end{flushright}
\vspace{0.5cm}
\begin{center}
{\Large\bf Induced Spin from the $ISO(2,1)$ Gauge Theory with the
Gravitational Chern-Simons Term}
\vskip 0.3cm
{\bf Jin-Ho Cho \footnote{e-mail : jhcho@phyb.snu.ac.kr}}

{\sl Department of Physics, KAIST\\
Taejon, 305-701, Korea}

\vskip 0.5cm
{\bf Hyuk-jae Lee \footnote{e-mail : lhjae@phya.yonsei.ac.kr}}

{\sl Department of  Physics Education,
Seoul National University\\
Seoul, 151-742, Korea}
\vskip 0.7cm
\end{center}
\setcounter{footnote}{0}
\begin{abstract}
In the context of $ISO(2,1)$ gauge theory, we consider $(2+1)$-dimensional
gravity with the gravitational Chern-Simons term (CST). This formulation
allows
the `exact' solution for the system coupled to a massive point particle
(which is not the case in the conventional Chern-Simons gravity). The
solution exhibits locally trivial structure even with the CST, although
still
shows globally nontrivialness such as the conical space and
the helical time structure. Since the solution is exact, we can say the CST
induces spin even for noncritical case of $\s+\al m\ne 0$.
\end{abstract}
\centerline{(To appear in Phys. Lett. B)}
\end{titlepage}
\newpage
\renewcommand{\thefootnote}{\arabic{footnote}}
\baselineskip=18pt

In the hope of unifying elementary forces including gravity, it has been
one of the most fascinating things in theoretical physics
to understand gravity in the context of gauge theory \cite{sci}.
There have been many approaches to this end starting from Kibble \cite{kib}.
However, due to the inhomogeneity and non-compactness of the Poincar\'{e}
group, it is still never trivial to construct a satisfactory theory that can
be viewed on an equal footing with other gauge theory. Recently,
Witten showed that at least in (2+1)-dimension, it is possible to write
Einstein
gravity in a completely analogous way with the usual Chern-Simons gauge
theory \cite{witten}. Further, motivated by the fact that the local
$ISO(2,1)$ symmetry is, on shell, equivalent to the usual diffeomorphism
\cite{townsend}, Grignani et al. constructed Poincar\'{e} gauge theory
coupled with a massive point particle, making use of `Poincar\'{e}
coordinates' \cite{nard,grig}.

The above Poincar\'e gauge gravity is based on a non-degenerate,
invariant quadratic form
$<J_a,\,P_b>=\e_{ab},<J_a,\,J_b>=<P_a,\,P_b>=0$ on the Lie algebra
$iso(2,1)$, where $P_a$ and $J_a$ are the generators satisfying
\be
[P_a,\,J_b]=\ep_{ab}{}^cP_c,\quad [J_a,\,J_b]=\ep_{ab}{}^cJ_c,
\quad [P_a,\,P_b]=0.
\ee
Note that the conventional Killing metric of $iso(2,1)$ is degenerate, so we
can't use it here.
Incorporating this quadratic form, the usual Chern-Simons Lagrangian for the
gauge connection $\CA=\o^aJ_a+e^aP_a$ leads to just
the the Einstein-Hilbert Lagrangian written in the vielbein notation.
\Be
\CL_{EH}&=&<\CA \wedge\!\!\!,(d\CA +\frac{2}{3}\CA\wedge\CA)>\nn\\
&=&2e^c\wedge(d\o_c+\frac{1}{2}\ep_{cab}\o^a\wedge\o^b).
\Ee

This raises a question; what about CST $\sim
\o^c\wedge(d\o_c+\frac{1}{3}\ep_{cab}\o^a\wedge\o^b)$ that was first
introduced by Deser et al. \cite{des} to give local dynamics to the
standard locally trivial (2+1)-dimensional Einstein gravity (hereafter, the
resulting Einstein gravity accompanied by CST is called Chern-Simons gravity
or CSG). We expect this
CST also to be formulated possibly in $ISO(2,1)$ gauge theory, just from
Chern-Simons Lagrangian, since the term is gauge invariant. Indeed
such can be done making use of general quadratic form obtained by taking the
Poincar\'{e} limit ($\l\goto 0$) for linear combination of
two types of de Sitter invariant quadratic forms which Witten proposed in
ref. \cite{witten}.
However, the same quadratic form is attainable more systematically, that is,
without resource to the above de Sitter invariant quadratic forms. We just
require those generators $P_a,J_b$ be the Killing vectors for the supposed
quadratic form. The resulting general quadratic form of $ISO(2,1)$ reads as
\be
<J_a,J_b>=\al\ \eta_{ab},\quad <J_a,P_b>=\eta_{ab},\quad <P_a,P_b>=0,
\ee
where $\al$ is some coefficient with the dimension of length.
We can easily check that this general quadratic form generates CST in
the Lagrangian.
\Be
\CL_{gauge}&=&<\CA \wedge\!\!\!,(d\CA +\frac{2}{3}\CA\wedge\CA)>\nn\\
&=&\al\,\o^c\wedge(d\o_c+\frac{1}{3}\ep_{cab}\o^a\wedge\o^b)
+2e^c\wedge(d\o_c+\frac{1}{2}\ep_{cab}\o^a\wedge\o^b).
\EE{lag}

Therefore, $ISO(2,1)$ gauge theory generically produces not only
Einstein-Hilbert term but also the
gravitational CST just from Chern-Simons Lagrangian. This seems to be in
contrast with the above mentioned CSG, where instead of
$ISO(2,1)$ symmetry, conformal symmetry is shown to be involved
\cite{witten}\cite{des}. The mystery lies in the fact that in CSG, they use
torsion
free condition as a constraint, thus the spin connection $\o^a{}_\m$ is a
functional of the fundamental variable $e^a{}_\m$, while in the $ISO(2,1)$
gauge theory, the torsion free condition comes out as an equation of motion
for the source free region and the
two connection components $\o^a$ and $e^a$
are independent variables.

Now we have another question; do those two
formalisms give the same geometry for a given spinless matter source,
as is usual in (3+1)-dimension. In this letter, we are to answer
this question, considering more general case,
a massive point source with its
intrinsic spin (the spinless case is straightforward). We first start from
the Lagrangian (\ref{lag}) and proceed to minimally couple
a massive spinning source. After fixing the
gauge degrees of freedom, we solve those equations to construct
the metric. Next, we discuss about the resulting geometry.

The usual Einstein-Cartan $SO(2,1)$ gauge theory deals with the `tangent
space' as the internal space and $SO(2,1)$ is the transformation group
acting on this internal space. However, to enlarge this gauge structure to
the inhomogeneous group $ISO(2,1)$, we consider, as the internal space, the
`affine tangent space', which is the tangent space with the freedom of the
specification of the origin \cite{kob}\cite{hennig}.
Since the affine coordinates $\f^a$
of the affine tangent space are $ISO(2,1)$ vectors (Poincar\'{e} vectors),
its covariant derivative is
\Be
\CD_\m\f^a=(\rd_{\m}\f +\CA_\m \f)^a
=\rd_\m \f^a +\ep^a_{\ bc}\o^b_{\ \m}\f^c
+e^a_{\ \m}.
\label{solder}
\Ee
It is to be understood that Latin indices $a=0,1,2$ specify the internal
affine space coordinates while Greek indices $\m=0,1,2$ specify
the space-time coordinates.
Making use of eq. (\ref{solder}), one can construct the physical length
that is invariant under coordinate-reparametrization
and gauge transformation,
\Be
dl^2 &=& \e_{ab} \CD_\m\f^a \CD_\n\f^b dx^\m dx^\n\nn \\
&=& g_{\m\n}dx^{\m}dx^{\n},
\label{mat}
\Ee
where $\CD_\m \f^a \equiv \CE_\m{}^a$ just plays the role of the soldering
form since the dimension of the affine
tangent space,
that of the tangent space and that of the space-time manifold
are all the same, that is, $det\,\CE\ne 0$.
This soldering form yields
the definition of the `physical' torsion two form \cite{nard}\cite{grig},
\Be
T^a\equiv\CD\CE^a=\CT^a +\ep^a{}_{bc}\CR^b\f^c,
\label{torsion}
\Ee
where
\Be
\CR^a=d\o^a+{1\over 2}\ep^a_{\ bc}\o^b\wedge\o^c,\quad
\CT^a =d e^a+\ep^a_{\ bc}\o^b\wedge e^c.
\Ee
Now we are to minimally couple a massive spinning particle to the system
(\ref{lag}). The Lagrangian for the interaction part is
\be
\CL_{int}=\int d\t \dot x^\m \d^3(x-x(\t))(j_a \o^a{}_\m+p_a e^a{}_\m),
\ee
where $j_a$ denotes the total angular momentum (orbital angular momentum
$l_a$ plus intrinsic spin angular momentum $s_a$) while $p_a$ represents
particle's momentum \cite{ger}\cite{ort}.
However in $(2+1)$-dimension,
the intrinsic spin $s^a$ being proportional to
the momentum $p^a$ with the proportionality coefficient $\s$ as the spin
scalar, because the second
Casimir invariant is the Pauli-Lubanski scalar $W=p^a j_a=p^a s_a$,
we may write $j^a$ generically as
\be
j^a =\ep^a_{\ bc}\f^b p^c +\frac{\s}{m} p^a,
\ee
where $m$ is particle mass. The whole action consequently amounts to
\Be
I&=&I_{particle}+I_{int}\nn\\
&&-\frac{1}{\k}\int d^3 x \ep^{\m\n\r}[\al\o_{a \m}(\rd_\n\o^a_{\ \r}
+\frac{1}{3}\ep^a_{\ bc}\o^b_{\ \n}\o^c_{\ \r})
+ e_{a \m}(\rd_\n\o^a_{\ \r}-\rd_\r\o^a_{\ \n}
+\ep^a_{\ bc}\o^b_{\ \n}\o^c_{\ \r})],\nn\\
\Ee
where $\k$ is the Einstein constant and we need not specify the particle
action $I_{particle}$ because we are only concerned about the geometry a
static massive spinning particle generates in $(2+1)$-dimension.

Straightforward variations of gauge connection components for the above
action result in the equations of motion,
\Be
\d\o^a_{\ \al}&\Goto&
\frac{1}{\k} \ep^{\al\r\g} (\al \CR^a_{\ \r\g}+\CT^a_{\ \r\g})
=\int d\t\d^3(x-x(\t))\dot x^{\al} j^a\nn\\
\d e^a_{\ \al}&\Goto& \frac{1}{\k}\ep^{\al\r\g} \CR^a_{\ \r\g}
= \int d\t\d^3(x-x(\t))\dot x^{\al} p^a,
\Ee
which together lead to
\be
\frac{1}{\k}\ep^{\al\r\g}\CT^a_{\ \r\g}=\int d\t\d^3(x-x(\t))\dot
x^{\al}[j^a-\al p^a].
\ee

In the above equation,
we note again that $\CT^a{}_{\m\n}$ is not the physical
torsion since the right hand side contains the orbital angular momentum; we
recall that the physical torsion is the intrinsic rotation generator,
conversely to say, the intrinsic spin is the source of the physical torsion.
{}From the above equations of motion, we proceed to the physical torsion
defined in (\ref{torsion}):
\Be
\frac{1}{\k}\ep^{\al\b\g}T^a{}_{\b\g}
&=&\frac{1}{\k}\ep^{\al\b\g}(\CT^a{}_{\b\g}
+\ep^{a}{}_{bc}\CR^b{}_{\b\g}\f^c)\nn\\
&=&\int d\t\d^3(x-x(\t))\dot{x}^\al[(\frac{\s}{m}-\al) p^a].
\Ee
Here, noting that $\al$ has the dimension of length, thus, $-\al p^a$ has
that of the augular momentum and it is the source of the physical torsion,
one may think of $-\al p^a$ as the spin induced from the gravitational CST.

Let us turn to solving the equations of motion to construct the metric
according to the eq. (\ref{mat}).
As a first step to this end, we first fix those gauges concerned with
the reparametrization symmetry and the internal $ISO(2,1)$ symmetry as
\Be
x^\m=\d^\m{}_a\f^a,\,\,\,\, \o^i{}_\m=0,\,\,\,i=1,2.
\Ee
It should be emphasized that only six components of the connection are fixed
because there are
six gauge degrees of freedom corresponding to the dimension
of $iso(2,1)$.
The above specific choice of gauge is for later convenience in
constructing the metric comparable with the usual Kerr solution
\cite{desera}-\cite{deserb}. Furthermore, without loss of
generality, we suppose that the rest particle is located at $(a^0,a^1,a^2)$
of the internal coordinates corresponding to the point at
$(\wt a^0, \wt a^1, \wt a^2)$ of the external space-time coordinates.
In this case, the momentum and the angular momentum read respectively as
\Be
p^a=(m,0,0),\qquad J^a=(0,-m a^2,m a^1)+(\s-\al m,0,0),
\Ee
where the first term of $J^a$ is the
orbital angular momentum while the second is the sum of the intrinsic
spin $\s$ and the component $-\al m$ induced from CST that is just the
consequence of the new quadratic form on $iso(2,1)$.

The above specifications of gauge and location lead the equations of motion
for the connection components to
\Be
& &\ep^{ij}\rd_j\o^0{}_0=0,\quad \ep^{ij}\rd_je^0{}_0=0,\nn\\
& &\ep^{ij}\rd_i\o^0{}_j=\frac{\k m}{2}\d^2(\vec{x}-\wt a),\nn\\
& &\ep^{ij}\rd_i e^0{}_j=\frac{\k (\s-\al m)}{2}\d^2(\vec{x}-\wt a),\nn \\
& &\ep^{jk}[\rd_j e^i{}_k +\ep^i{}_l \o^0{}_j e^l{}_k]=-\frac{\k m}{2}
\ep^i{}_k a^k \d^2(\vec{x}-\wt a),\nn\\
& &\ep^{ij}[\rd_je^k{}_0+\ep^k{}_l(\o^0{}_je^l{}_0-\o^0{}_0e^l{}_j)]=0.
\Ee
With appropriate boundary conditions and the stationary condition, one
can achieve the following simplified solutions.
\Be
\o^0{}_0 &=& e^0{}_0\,\,=\,\,e^i{}_0\,\,=\,\,0,\nn\\
\o^0{}_j &=& \frac{\k m}{4\pi} \frac{\ep_{ij}(\f^i
-a^i)}{|\vec{\f}-\vec{a}|^2},\nn\\
e^0{}_j &=& \frac{\k (\s-\al m)}{4\pi} \frac{\ep_{ij}(\f^i
-a^i)}{|\vec{\f}-\vec{a}|^2},\nn\\
e^i{}_j &=& \d^i{}_j \frac{A^{\frac{\k
m}{4\pi}}}{|\vec{\f}-\vec{a}|^{\frac{\k m}{4\pi}}}
-\d^i{}_j+\frac{\k
m}{4\pi}\frac{\ep_{jk}\ep^i{}_l\f^l(\f^k-a^k)}{|\vec{\f}-\vec{a}|^2},
\Ee
where $A$ is some constant with the dimension of length and is introduced to
adjust the dimensionality of the components $e^i{}_j$.
Inserting these solutions into (\ref{mat}) results in the following
familiar form of length element.
\Be
ds^2&=&-[d\f^0+\frac{\k (\s-\al m)}{4\pi}\frac{\ep_{ij}(\f^i - a^i)}
{|\vec{\f}-\vec{a}|^2}d\f^j]^2+\frac{A^\frac{\k m}{2\pi}}
{|\vec{\f}-\vec{a}|^{\frac{\k m}{2\pi}}}[(d\f^1)^2+(d\f^2)^2]\nn\\
&=&-(dt+\frac{\k (\s-\al m)}{4\pi}d\th)^2
+\frac{A^\frac{\k m}{2\pi}}
{|\vec{x}-\wt a|^\frac{\k m}{2\pi}}(dr^2+r^2d\th^2).
\Ee
This result is very similar to the metric
obtained by Deser \cite{deserb} and Linet \cite{linet} from
Einstein-Cotton equation in the linearized approximation.
However, we should note that the signature of the induced spin $(-\al m)$
is different from theirs $(\al m)$. Moreover in CSG, it was shown
to be impossible to get any exact
solution with Deser's asymptotic limit except when $\s+\al m=0$ \cite{cle}.
Consequently, it is remarkable that the $ISO(2,1)$ gauge formulation allows
the above exact solution without any specific condition on the induced spin.
We can say in the gauge formulation, the induced spin manifests itself from
the gravitational CST independent of the intrinsic  spin of the particle.

We conclude this letter with some remarks on our results.
The solution exhibits flat structure even in the presence of CST. This is in
contrast with other Chern-Simons gauge theory.
This difference possibly stems
from the fact that in $ISO(2,1)$ gauge theory,
both the Einstein-Hilbert term
and CST involve one derivative,
whereas in other conventional Chern-Simons gauge theory, since the Maxwell
term involves two derivatives, the one derivative
involved Chern-Simons term
effectively becomes mass term. Therefore in $ISO(2,1)$ gauge theory, CST
produces no topologically massive mode. Gauge formulation of
topologically massive gravity can be achieved by introducing an extra
Lagrange multiplier field to put the torsion free condition into the
Lagrangian as a constraint. This was dealt with in refs. \cite{xiang}
and \cite{carlip}.

However as
in (2+1)-dimensional Einstein gravity, the global structure of the
resulting geometry is still nontrivial; the conical space structure of
the deficit angle $\k m/2$ and the time helical structure (time jumps by
$\k(\s-\al m)/2$ for a $2\pi$ increase of $\th$) are apparent here.
The momentum is the source of the
translation part of the curvature and produces the conical space structure
while the total angular momentum, together
with the induced term $-\al\,p^a$ becomes
the source of the Lorentz part of the curvature and gives the time helical
structure.

We lastly note that in contrast with $(3+1)$-dimensional case, the
Einstein formulation (torsion free condition assumed) and gauge formulation
result in totally different pictures in this case.
For the spinless case in $(3+1)$-dimension, the two formulations give
the same result.
However in $ISO(2,1)$ gauge theory, due to the spin induced
from CST, torsion gives the time helical structure in the absence of any
spinning source, while in CSG, the same term gives topologically massive
mode and produces locally nontrivial geometry.

\vskip 2truecm
\noindent
{\large\bf Acknowledgement}\\
The authors were supported by the Korea Science and
Engineering Foundation.

\end{document}